\documentclass[aps,prd,onecolumn,12pt,groupedaddress,nofootinbib]{revtex4}

\usepackage{amsmath}
\usepackage{amssymb}
\usepackage{graphicx}

\begin{document}

\title{Inspecting absorption in the spectra of extra-galactic gamma-ray sources for insight into Lorentz invariance violation}
\author{Uri Jacob}
\email{uriyada@phys.huji.ac.il}
\author{Tsvi Piran}
\affiliation{Racah Institute of Physics, The Hebrew University, Jerusalem, Israel}

\begin{abstract}
We examine what the absorbed spectra of extra-galactic TeV gamma-ray sources, such as blazars, would look like in the presence of Lorentz invariance violation (LIV). Pair-production with the extra-galactic background light modifies the observed spectra of such sources, and we show that a violation of Lorentz invariance would generically have a dramatic effect on this absorption feature. Inspecting this effect, an experimental task likely practical in the near future, can provide unique insight on the possibility of LIV.
\end{abstract}


\maketitle

\section{Introduction}
Quantum-gravity theories predict in general the breakdown of familiar physics when approaching the Planck energy scale ($E_{pl}=\sqrt{\hbar c^5/G}\sim1.2\times10^{28}$ eV). A violation of Lorentz invariance arises naturally in various such theories (for a review see e.g. \cite{QG1,QG2}). Though the possible effects of LIV are not currently measurable on Earth, there is vast interest in exploring LIV, as it could direct us to the fundamental physical theory. We would like to obtain experimental evidence for a LIV scenario or, to the contrary, evidence strengthening the validity of Lorentz invariance and constraining the possibility of LIV. \par

We consider here a general approach that involves the breakdown of Lorentz symmetry at some very high energy scale, $\xi E_{pl}$. At lower energy scales small deviations from Lorentz invariance may be felt. This kind of LIV approach is inspired by the theoretical work on quantum-gravity, but the phenomenology may of course be considered independently of quantum-gravity. The basic phenomenology of LIV can be described by a modified dispersion relation. When examining particles with $E\ll\xi E_{pl}$, we may regard only the leading order correction to Lorentz invariance and generically (for ultra-relativistic particles) write the dispersion relation as:
\begin{equation}\label{LIDform}
E^2-p^2c^2-m^2c^4\simeq\pm E^2\left(\frac{E}{\xi_n E_{pl}}\right)^n,
\end{equation}
where the dimensionless parameter, $\xi_n$, and the order of the leading correction, $n$, are model parameters (which can be universal or be given separately for each particle type, depending on the framework). $c$, the conventional speed of light constant, remains only the speed of low-energy massless particles, and we put hereafter $c=1$. We notice, assuming that the standard relation $v=dE/dp$ holds, that the $+$ and $-$ signs account for energy-dependent superluminal and infraluminal motions. \par
This dispersion relation points us to the most direct method of testing LIV - inspecting particles' speeds. As the effects of LIV may become substantial for high-energy particles, there has been much work recently on testing LIV by means of time-of-flight measurements of astronomical particles. Time-of-flight analyses provide the most generic method of testing LIV and have been performed for different cosmological gamma-ray sources. No signatures of LIV have been detected, yet these observations were at most sensitive to effects of LIV with a scale two orders of magnitude below the Planck scale \cite{AGN-TOF,Crab-TOF,Boggs,MTY,Ellis2,CosmoDel,Ellis3}. Time-of-flight measurements of cosmological neutrinos are expected to be more sensitive \cite{ourNature}. \par

Another observational option for examining LIV is to inspect the thresholds of high-energy particle reactions, as these might be shifted as a result of a LIV alteration of the transformations between reference-frames. This test is not strictly kinematic and so less generic. Nevertheless, the thresholds can be derived directly from the deformed dispersion relation (with the sole assumption being that it represents the only modification of standard physics). Regardless of the specific theoretical details and the particle dynamics within a given framework, one can expect that theories where the Lorentz symmetry is broken at extreme energies would be manifested in the low-energy regime by an approximated dispersion relation of the form (\ref{LIDform}). Thus, this method for gaining insight on LIV, dealing with particle interactions, is not model-specific and is also an efficient approach. \par

The cutoff in the spectrum of ultra high energy cosmic rays, the so-called GZK effect \cite{G,ZK}, which in the past failed to be observed as expected, created an apparent paradox and originally motivated the consideration of LIV as a possible solution \cite{GZK-LIV1,CG,LIDparadox}. LIV could have pushed upward the threshold energy of the reaction between the cosmic rays and the cosmic microwave background (CMB) photons. The GZK cutoff has more recently been detected by the HiRes observatory \cite{HiRes} and then by Auger with high significance \cite{Auger}. Turning the previous argument around, the agreement of the GZK threshold with the classical expectation can be used \cite{ThreshConst} to place a very severe bound on LIV well above the Planck scale. However, some theoretical frameworks predict a priori that LIV does not affect cosmic rays (as any charged particles), making the above bound less generally valid \cite{EquivViol,EllisBrief,EllisNew}. \par

Considering that in some LIV frameworks only photons deviate from classical behavior, phenomena involving just photons are the most suitable LIV tests. Specifically, it has been suggested \cite{TeVgamma-LID1,TeVgamma-LID2,Aloisio,LIDparadox} that LIV could affect the cosmological interaction between gamma-rays and infra-red background photons. We proceed to demonstrate here the feasibility of testing LIV by investigating absorption in the TeV spectra of gamma-ray sources. We show that this can supply unique evidence for LIV (or rule out its existence) at the Planck scale level. \par

The paper is organized as follows. In Sec. \ref{sec:Thresholds} we review the effect of LIV on the pair-production interaction threshold. Section \ref{sec:EBLabsorption} describes the extra-galactic background field and its impact on the optical depth and observation of gamma-ray sources. In Sec. \ref{sec:LIVabsorption} we explain and demonstrate how LIV would affect the absorption features of cosmological TeV gamma sources. We discuss our results and the practicality of gaining valuable insight on LIV in Sec. \ref{sec:Discussion}.

\section{Deformed interaction thresholds} \label{sec:Thresholds}
Numerous particle processes, such as the two mentioned above, only take place beyond a threshold, required for the conservation of energy-momentum and classically determined by Lorentz-transforming to the center-of-mass reference frame. The photopion production, responsible for the GZK effect, does not occur if the energy in the center-of-mass frame is not sufficient for the creation of a pion. The pair creation interaction, in which we are interested, can happen if in the center-of-mass frame the photons have a total energy larger than the rest mass of the electron-positron pair. Within LIV the transformation to the center-of-mass frame is deformed and the threshold energy in the laboratory frame can be shifted. Since in the context of a generic LIV approach we do not have global laws for transformation between reference frames, we must work fully in the laboratory frame. The interaction threshold can still be determined by equating the energy and momentum of the particles before and after the reaction, using the particles' dispersion relations and requiring that the solution to the equations is real at the threshold or above and imaginary below it (see \cite{LIDparadox}). The single assumption made here is that standard energy-momentum conservation is maintained in particle reactions (note that this assumption does not hold in Doubly Relativity models \cite{DoublyPhen}). \par

The classical prediction of Special Relativity is that the threshold energy of a soft photon for pair-production by head-on collision with a gamma-ray of energy $E_\gamma$  is $m_e^2/E_\gamma$, where $m_e$ is the electron rest mass. Within our LIV phenomenology we use (\ref{LIDform}) and arrive, as explained, at the modified pair-production threshold:
\begin{equation}
\epsilon_{thr}=m_e^2/E_\gamma+\frac{1}{4}\left(\frac{E_\gamma}{\xi_n E_{pl}}\right)^nE_\gamma,
\end{equation}
if the deformed dispersion relation is only applied to photons, or
\begin{equation}\label{LIVthreshold}
\epsilon_{thr}=m_e^2/E_\gamma+\frac{1}{4}(1-2^{-n})\left(\frac{E_\gamma}{\xi_n E_{pl}}\right)^nE_\gamma,
\end{equation}
if LIV affects all particles equally. We want our results to be relevant for theories where the Lorentz-violating physics applies only to photons (and not electrons) as well as for theories where the LIV correction is universal (natural when the correction is due to the small-scale structure of spacetime and the equivalence principle holds). However, we find, as far as the interaction of gamma-rays with background photons is concerned, a similar LIV effect. The application of LIV to electrons accounts for the reduction of the LIV term in (\ref{LIVthreshold}) by a maximal factor of $2$. This is equivalent to modifying $\xi$ by such a factor, and we show subsequently that our results depend very weakly on the choice of the symmetry breaking scale. We, therefore, use for the rest of our study the threshold energy expression of (\ref{LIVthreshold}), as a more conservative estimate of the LIV effect, and our findings will be generically valid. The threshold expression of (\ref{LIVthreshold}) includes the leading order correction in the small factors $\frac{E_\gamma}{\xi_n E_{pl}}$ and $\frac{m_e}{E_\gamma}$ and accounts for a LIV scenario with subluminal motion [`$-$' sign in Eq. (\ref{LIDform})]. Such a LIV scenario, which produces an upward shift of interaction threshold energies, is more probable theoretically and observationally, and we proceed to study it. \footnote{A LIV scenario with superluminal motion would have the correction term in (\ref{LIVthreshold}) negative. The interaction threshold energy would be lowered, approaching zero above $E^*$. This would imply that gamma-rays of energy $\sim E^*$ can interact with the bright CMB field, resulting in a cutoff of all extra-galactic gamma-ray sources around $E^*$. This LIV scenario therefore has a very clear signature, and it is unlikely with current observations.} \par

Inspecting (\ref{LIVthreshold}) we find that in the context of LIV there is a critical energy, which we denote $E^*$, for which $\epsilon_{thr}$ becomes minimal. This new phenomenon at $E^*$ is a key ingredient in our approach. The critical energy is straightforwardly derived by setting $\epsilon_{thr}'(E_\gamma)=0$. The general expression is:
\begin{equation}\label{EStar}
E^*=\left(\dfrac{4m_e^2(\xi_n E_{pl})^n}{(n+1)(1-2^{-n})}\right)^\frac{1}{n+2}.
\end{equation}
It is apparent that the dependence on the symmetry breaking scale is weak, especially for $n=1,2$. With $n=1$ varying $\xi_1$ by two orders of magnitude results only in changing $E^*$ from $11$ TeV for $\xi_1=0.1$ to $50$ TeV for $\xi_1=10$. We happen to be lucky that this is a convenient energy range and for $\xi_1=1$ (the symmetry breaking scale being the Planck energy) the critical $E^*$ is at the examinable energy of $23$ TeV. For $n=2$ $E^*\sim10^{5}$ TeV with the symmetry breaking around the Planck scale. Scenarios with $n\geqslant2$ are, therefore, clearly unapproachable. We carry on demonstrating the applicability of our method for the investigation of the theoretically motivated $n=1$, $\xi\sim1$ LIV scenario. \par

\begin{figure}
   \includegraphics[width=16cm,clip=true]{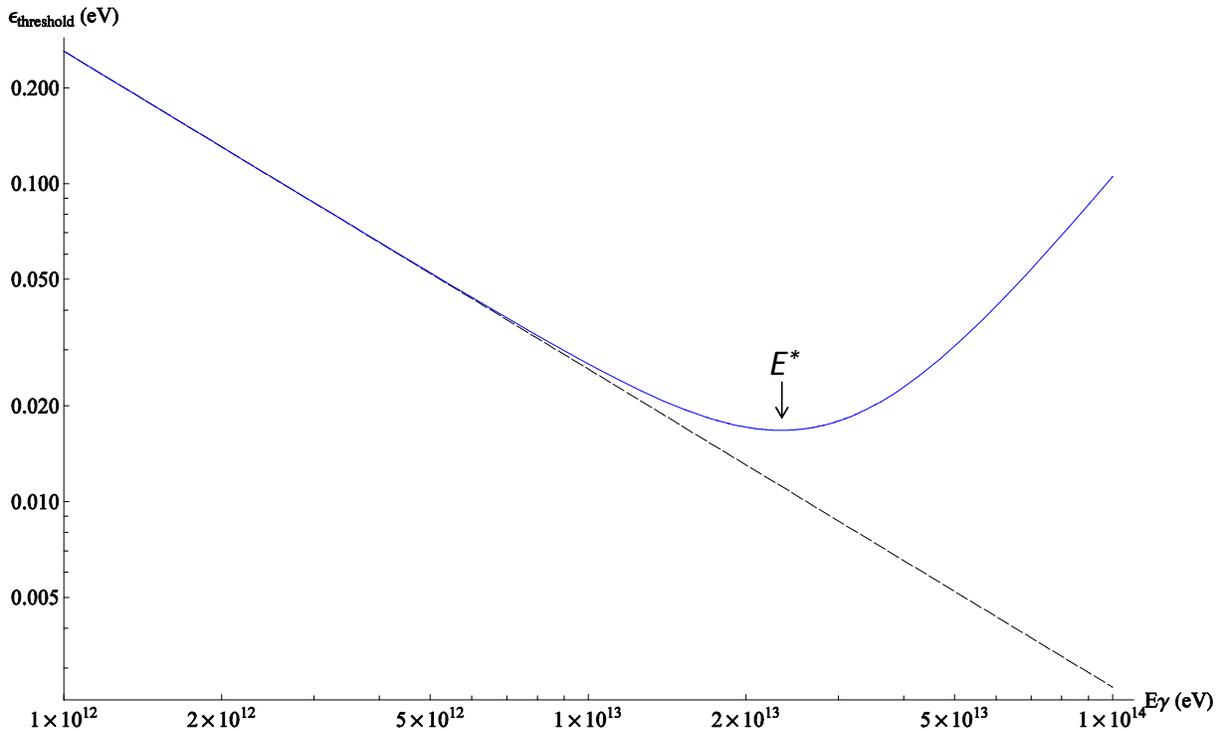}
   \caption{The soft photon pair-production threshold energy as a function of the interacting hard photon energy - according to the classical calculation of Special Relativity (dashed black line) and with $n=1$, $\xi=1$ LIV (blue curve).}
   \label{fig:Thresholds}
\end{figure}

We present in Fig. \ref{fig:Thresholds} the shift of pair-production threshold energies caused by LIV.
It is apparent that at low gamma-ray energies the effect of LIV is negligible, while it becomes qualitatively significant when approaching $E^*$. For $E^*$ the soft photon threshold energy reaches a global minimum and then increases with increasing gamma-ray energy (as opposed to the systematic decrease in the Lorentz invariant case). We notice that some soft photons (at energies below $\sim 0.02$ eV with the LIV parameters here) never interact with any hard photons, as they are below the pair-production threshold for every gamma-ray energy. For every soft photon that does interact with gamma-rays above a lower threshold, there is also an upper gamma energy threshold above which the interaction does not take place. We explore the implications of these peculiar high-energy interaction properties in the following.

\section{The extra-galactic background radiation and cosmological gamma-ray sources} \label{sec:EBLabsorption}
The extra-galactic medium contains uniform background radiation fields in several spectral ranges. The part of the spectrum that consists of the infra-red, optical and ultra-violet backgrounds is commonly referred to as the extra-galactic background light (EBL). The EBL, which shines at energies just above the familiar CMB, is thought to consist of the accumulated redshifted thermal emission of stars and dust throughout the history of the universe. Yet, as opposed to the well-resolved CMB, this background is difficult to determine, as direct measurements are problematic, owing to the bright galactic and solar system foregrounds present. Certain estimates and limits on the EBL spectrum exist. In the infra-red range, of highest relevance to our study, they arise mostly from COBE (Cosmic Background Explorer) and IRTS (Infrared Telescope in Space) observations and from galaxy counts. Still, there are significant uncertainties regarding the EBL spectrum, and theoretical estimates are very cosmological-evolution model-dependent \cite{COBE,GalCount,IRbackRev,IRTS}. \par

Gamma-rays from cosmological sources are attenuated by pair-production interaction with the EBL, so a source appears on Earth with a different spectrum than its emitted one. The observed gamma-ray spectrum from a source at redshift $z_s$ is given by $I_{obs}[E_{\gamma}]=I_{int}[E_{\gamma}(1+z_s)]e^{-\tau(E_{\gamma},z_s)}$, where $I_{int}$ is the intrinsic source spectrum and $\tau$ is the respective optical depth. The calculation of the optical depth uses the pair-production cross-section \cite{PPcross} and is determined by the spectral energy distribution of the EBL and the length of the inter-galactic path over which the gamma-rays propagate:
\begin{equation}\label{OpticalDepth}
\tau(E_{\gamma},z_s)=\int_0^{z_s}\!dl(z) \int_0^\infty\!d\epsilon\; \sigma_{\gamma\gamma}\left[E_{\gamma}(1+z),\epsilon(1+z)\right]n(\epsilon)(1+z)^2,
\end{equation}
where $dl$ is the proper line element (we use standard cosmology parameters), $\sigma_{\gamma\gamma}$ is the cross-section for $\gamma\gamma\rightarrow e^-e^+$ in an isotropic field of background photons and $n(\epsilon)$ is the EBL photon number distribution at present. We employ a simplistic model for the cosmological evolution, where $n_z[\epsilon(1+z)]=n(\epsilon)(1+z)$. This is negligible in our study which is concerned with low redshift sources. \par

Over recent years several extra-galactic sources (mostly blazars) have been observed by imaging air Cherenkov telescopes (IACTs) in the TeV gamma-ray range. The spectra of some observed blazars seem to be harder than anticipated, considering the expected EBL absorption. However, this expectation is ambiguous because of the uncertainty in the EBL spectrum. Several studies have, thus, attempted to extract a more reasonable EBL spectrum out of the observed blazars spectra. This is also somewhat problematic, since our knowledge of the intrinsic sources spectra is limited \cite{IRabsorption,BlazarSpectra,EBLblazarConst,EBL-blazars}. Other studies have further suggested that the harder than anticipated spectra could be due to LIV effects \cite{LIDparadox,TeVgamma-LID2}.
Given the uncertainties it seems unreasonable to justify an exotic explanation such as LIV effects in the context of such studies. We use the exceptional structure of the Lorentz-violating interaction threshold to suggest in this study a simple analysis of observational data that can give a clear sign of LIV or conclusively rule it out regardless of EBL specifics.

\section{Absorbed gamma-ray sources in the presence of LIV} \label{sec:LIVabsorption}
The pair-production reaction's cross-section is concentrated in a narrow energy range above the threshold (peaking at $\sim4$ times the threshold energy). As classically the soft photon interaction threshold is a monotonous function of the gamma-ray energy, given that lower energy background photons are more abundant than higher energy ones [the spectrum is steeper than $n(\epsilon)=\epsilon^{-1}$ in the relevant range], the optical depth becomes monotonously larger for higher energy gamma-rays. For every cosmological source distance there is a gamma-ray cutoff energy at which the optical depth becomes unity, and around it the observed spectrum should show exponential suppression. LIV would change this behavior drastically. In the presence of LIV there is a certain gamma-ray energy, $E^*$ (see Sec. \ref{sec:Thresholds}), above which higher energy gamma-rays require higher energy EBL photons for the pair-production interaction. For any cosmological source the optical depth would thus be maximal at $E^*$, and at higher energies it would rapidly decrease. \par

In the classical calculation of the optical depth (see \cite{PPcross}) the pair-production cross-section is determined by the soft photon's energy relative to the threshold for interaction with a hard photon of $E_{\gamma}$, i.e. $\sigma_{\gamma\gamma}[E_{\gamma},\epsilon]=\sigma_{\gamma\gamma}[\epsilon/\epsilon_{thr}(E_{\gamma})]$. To proceed within a generic LIV approach we assume that the modified cross-section, $\sigma_{\gamma\gamma,LIV}$, has the same dependence on $\epsilon/\epsilon_{thr}$. The cross-section is merely shifted as we now evaluate it by regarding the soft photon's energy relative to the LIV shifted pair-production threshold - we use the $\epsilon_{thr}$ of (\ref{LIVthreshold}) to evaluate $\epsilon/\epsilon_{thr}(E_{\gamma})$. When $\epsilon_{thr,LIV}(E)=\epsilon_{thr,classical}(E_0)$, we have $\sigma_{\gamma\gamma,LIV}[E,\epsilon]=\sigma_{\gamma\gamma,classical}[E_0,\epsilon]$, and we can put in (\ref{OpticalDepth}) $\sigma_{\gamma\gamma,LIV}$ instead of $\sigma_{\gamma\gamma,classical}$. This is natural within any relativistic framework, but since the LIV framework might have a preferred inertial frame, the implied assumption is that modifications to the shape of the pair-production cross-section would constitute lower order corrections. With this basic assumption we find at any energy above $E^*$ an optical depth identical to the optical depth at some energy below $E^*$ for which the pair-production threshold is the same. This "symmetric" behavior is the key to our proposed LIV test. \par

\begin{figure}
   \includegraphics[width=16cm,clip=true]{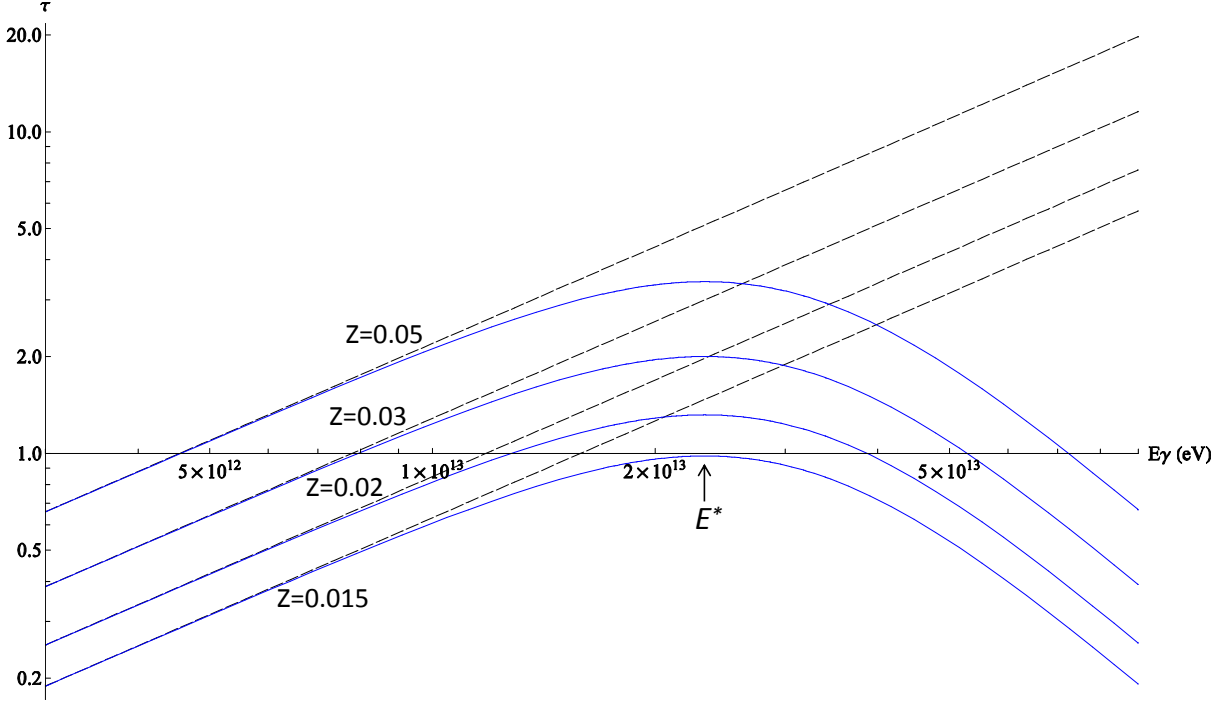}
   \caption{The optical depth for extra-galactic sources at sample redshifts as a function of gamma-ray energy. We use a simplified EBL spectrum (see text) and present the results with $n=1$, $\xi=1$ LIV (blue curves) and without LIV (dashed black lines).}
   \label{fig:OpticalDepths}
\end{figure}

The effect of the modified pair-production threshold on the optical depth of gamma-rays can be seen in Fig. \ref{fig:OpticalDepths}.
We used for simplicity an EBL spectrum of $n(\epsilon)=10^{-3}\epsilon^{-2}$ [eV$^{-1}$cm$^{-3}$] (approximately the average energy dependence of the EBL at the relevant infra-red energies), as the LIV features examined are not sensitive to this choice. We notice that for sources at any cosmological distance the optical depth is maximal at $E^*=23$ TeV. For some sources the optical depth reaches unity at a lower energy, increases until $23$ TeV and then decreases back to unity and below at higher energies. More nearby sources have optical depths that never reach unity, as they have $\tau(23$TeV$)<1$. While the detailed shape of the optical depth at gamma-ray energies up to $E^*$ depends on the detailed densities of the EBL spectrum, the optical depth above $E^*$ does not depend on additional background spectrum data. At any energy above $E^*$ the optical depth is the product of interaction with the same part of the EBL spectrum that produces the optical depth at some energy below $E^*$ (see Fig. \ref{fig:Thresholds}). The correspondence between values on both sides of $E^*$ is strictly dependent on the LIV parameters. This means that whatever the details of the EBL and the optical depth below $E^*$ are, with the LIV model we can foretell what the optical depth above $E^*$ will look like. For every source redshift from which a cutoff of the gamma-ray emission by the EBL absorption is observed, we can foretell the energy of re-emergence, where the optical depth decreases below unity again. \par

\begin{figure}
   \includegraphics[width=16cm,clip=true]{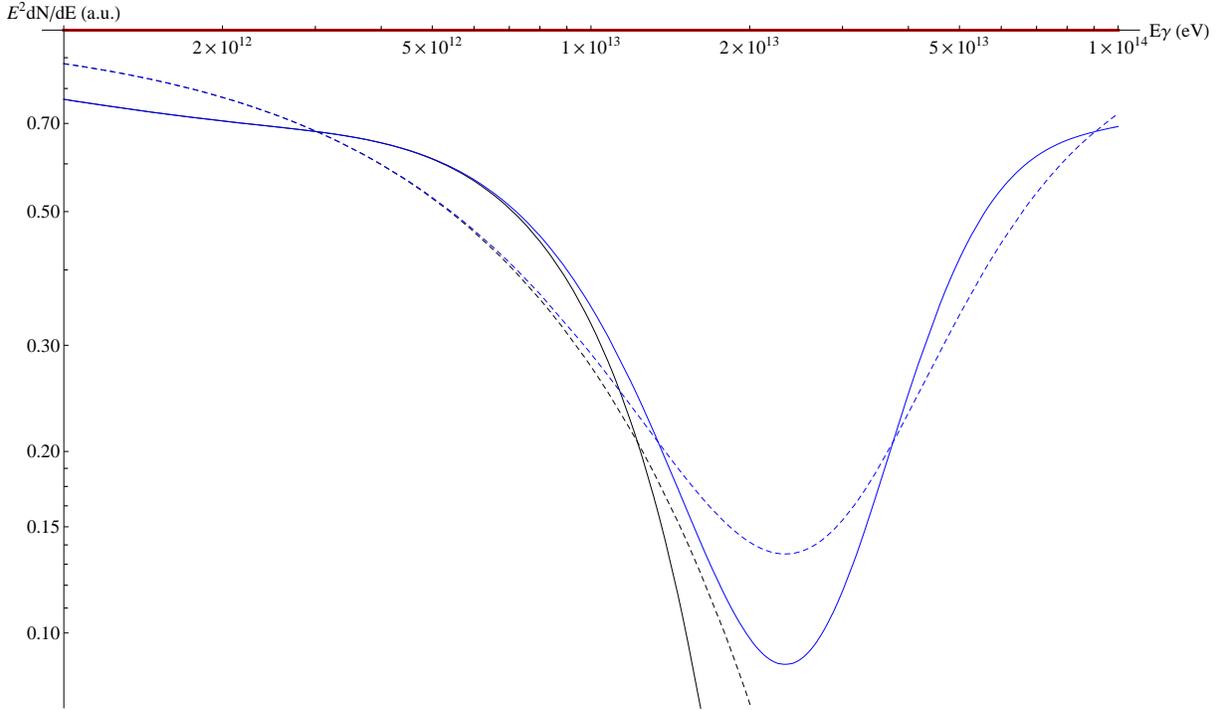}
   \caption{A sample blazar's spectrum on Earth in the presence of LIV at the Planck scale. We assume a source at $z=0.03$ with a simple intrinsic spectrum (see text) marked by the thick red line. The observed blazar spectrum is presented with the simplified EBL spectrum of Fig. \ref{fig:OpticalDepths} (dashed curves) as well as with a more realistic EBL spectrum (solid curves, see text). Blue curves describe the extra-galactic absorption within LIV, whereas black curves describe classically expected absorption.}
   \label{fig:BlazarSpectrum}
\end{figure}

Figure \ref{fig:BlazarSpectrum} shows how LIV would affect a blazar's observed spectrum.
We demonstrate this with a more realistic realization of the EBL spectrum, $\nu I_\nu=10$ nW/m$^2$/sr below $1.5\mu m$, going as $\lambda^{-1}$ up to $10\mu m$ and as $\lambda^1$ above (the EBL wavelengths relevant for attenuation of TeV gamma-rays are $\sim1$-$100\mu m$). Of course this is yet one realization out of many possible (see \cite{IRbackRev}). Our intention is simply to display the manifestation of the general effect. The figure also depicts the observed blazar's spectrum anticipated when considering the simplified EBL spectrum presented earlier. We notice that the principal LIV effects are the same. The sample blazar's intrinsic spectrum is chosen to be a power-law with a spectral index $(dN/dE\propto E^{-\alpha})$ of $\alpha=2$. This is a likely approximate spectral index for blazar emission in the TeV range \cite{BlazarSpectra,IntrinsicEBL}. We may expect to be able to estimate the intrinsic spectrum of a TeV gamma-ray source from lower energy measurements, where the effect of the extra-galactic attenuation is negligible. Of course the source's intrinsic spectral shape may vary and there may be an intrinsic break in the spectrum, but the accumulation of observations of different TeV gamma-ray sources will help us separate intrinsic source features from those induced by the EBL interaction (see discussion later). \par

\begin{figure}
   \includegraphics[width=16cm,clip=true]{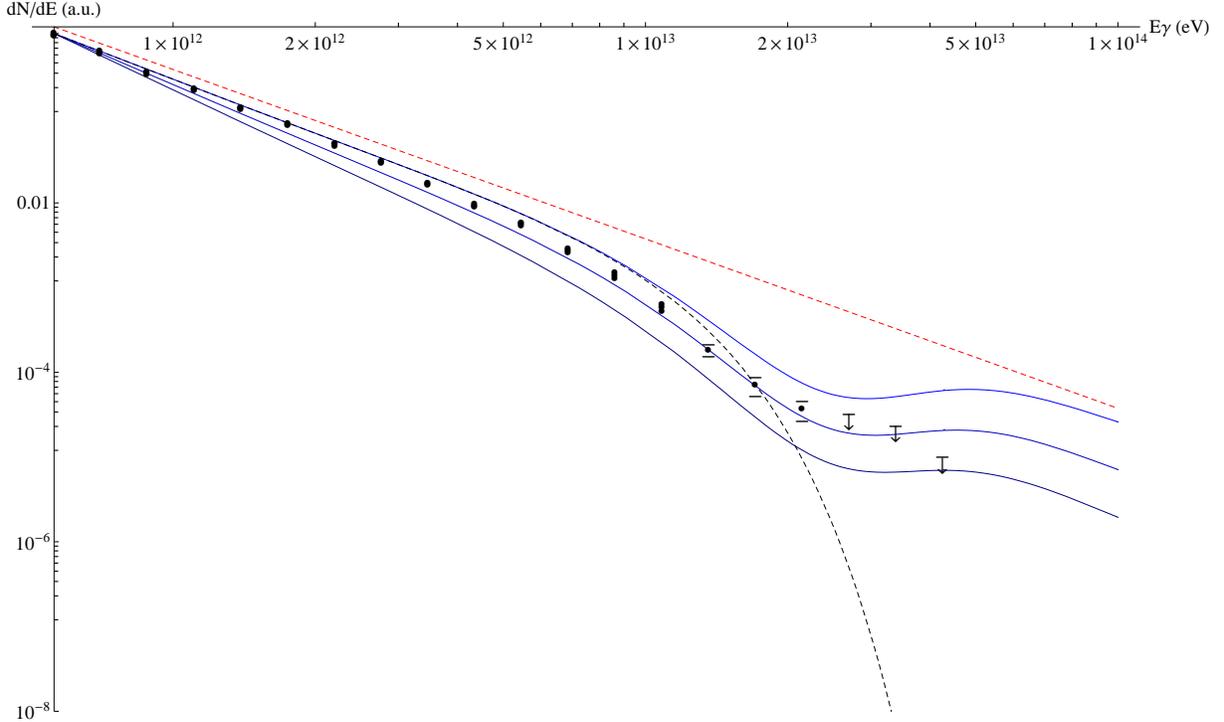}
   \caption{The expected photon fluxes from different sample blazars in the presence of LIV at the Planck scale. We examine three sources at a redshift $z=0.03$ with intrinsic power-law spectra, $dN/dE_{int}\propto E^{-\alpha}$. The blue solid curves describe the observed spectra of the sources with $\alpha=2,2.25,2.5$ (from top to bottom), considering LIV absorption with the same EBL realization as the solid curves in Fig. \ref{fig:BlazarSpectrum}. The dashed black curve describes the observed spectrum of the $\alpha=2$ blazar with no LIV. The dashed red line corresponds to the intrinsic spectrum $dN/dE\propto E^{-2}$. The normalized time-averaged fluxes of Mkn-501, measured during the 1997 outburst \cite{Mkn501burst}, are presented as black points with statistical errors. At higher energies we have $2\sigma$ upper flux limits (systematic errors are not included).}
   \label{fig:Detection}
\end{figure}

We are also faced with the consideration of whether enough photons will be detected by the gamma-ray observatories at energies above $E^*$, so that a source's spectral shape at these energies can be determined. Fortunately, the LIV scenario allows us in principle to observe nearby blazars at much higher multi-TeV energies than in the classical case, as the optical depth decreases instead of rapidly increasing. We demonstrate this in Fig. \ref{fig:Detection} with the expected photon fluxes from sources with different intrinsic spectra.
Mkn-501, a powerful blazar at a redshift $z=0.034$, has been observed up to $\sim20$ TeV and was found to have a spectrum that can be interpreted as a power-law of $\alpha\sim2$ with an exponential cutoff due to EBL extinction \cite{Mkn501burst}. As the $n=1$, $\xi=1$ LIV model does not have a significant effect on the optical depth below $20$ TeV, this observed spectrum can also be easily fit to absorption with LIV (where we may also use the freedom of the EBL realization). Above $\sim20$ TeV, however, LIV would qualitatively change the observed spectrum. The HEGRA observation of Mkn-501 during the extraordinarily active period in 1997 produced upper flux limits up to $\sim40$ TeV (see Fig. \ref{fig:Detection}). These upper limits are not stringent enough to rule out the LIV scenario - they seem to exclude an intrinsic power-law of $\alpha\lesssim2$ for Mkn-501, but the data may still be reasonably fit to LIV with $\alpha\sim2.25$ (taking into account the systematic error arising from the $15\%$ uncertainty on the absolute energy scale improves this fit). However, as apparent from Fig. \ref{fig:Detection}, we are on the verge of obtaining significant insight on LIV with symmetry breaking at the Planck scale. An observation of Mkn-501 at slightly higher energies would clearly distinguish between a LIV and a no-LIV scenario. Stronger upper limits at the high energies would tend to rule out LIV. Such observations could be possible with improved observatories or with another high-activity period of the source.

We see in Fig. \ref{fig:Detection} that while in the classical scenario the extra-galactic extinction becomes so strong that we cannot observe gamma rays much above $20$ TeV even with a flatter intrinsic spectrum, within the LIV scenario it is likely that we will be able to detect photons at higher energies regardless of the exact intrinsic spectral index. As an example, if such a nearby blazar's intrinsic TeV spectrum is $dN/dE\propto E^{-2}$, then, given LIV, the number of photons expected to arrive from it at the range of $20-40$ TeV is only a factor $\sim3$ smaller than the number at $10-20$ TeV. Furthermore, at $40-80$ TeV the number of expected photons is a factor $\sim2$ smaller than at $10-20$ TeV. Hence, if the LIV model holds, it is a realistic expectation to detect its signature with future observations. Figure \ref{fig:Detection} also shows that this signature is quite independent of the intrinsic power-law shape. All in all, although we do not know the specifics of the blazars' intrinsic spectra or the EBL spectrum, we see that LIV would have such a dramatic and unique effect on the observed gamma-ray spectra that we can expect to notice it and single it out from other effects. \par

Having demonstrated the detectability of the LIV effect for the scenario of $\xi_1=1$, we should note that if the symmetry breaking scale is much larger, the discussed LIV test will most likely not be feasible with the current generation of IACTs. While the dependence of the critical LIV energy on the symmetry breaking scale is weak (see Sec. \ref{sec:Thresholds}), for $\xi_1=10$ $E^*$ reaches $\sim50$ TeV. For a blazar such as those described in Fig. \ref{fig:Detection}, it is apparent that the photons reaching Earth will be drastically diluted at energies below $50$ TeV, so that the source will be completely unobservable at the energies where the effect of LIV would become substantial. Only if more nearby powerful extra-galactic sources are observed (see Fig. \ref{fig:OpticalDepths}), the gamma-rays at these high energies may be detected so that such weaker LIV scenarios can be examined. In any way, the examination of LIV scenarios with symmetry breaking around the Planck scale will be an important achievement, as such scenarios are commonly predicted by theories and no generic test for them has yet been performed.

\section{Discussion} \label{sec:Discussion}
We show here that TeV blazar observations could distinguish LIV effects and provide unambiguous results. We can see in Fig. \ref{fig:OpticalDepths} that a source which displays an exponential cutoff due to EBL absorption in the presence of LIV at any energy below $E^*$ would show a re-emergence of gamma-rays at another energy above $E^*$. The observation of such a strange phenomenon could clearly serve as a starting point leading us to LIV. Yet, because of the exotic nature of the LIV conjecture this would still not be convincing enough. However, if we observe the very-high-energy emission of a few different blazars, the ensemble of observations from different redshifts could serve to verify the LIV hypothesis. The different absorbed spectra must, of course, obey the same LIV parameters, hence we can assess all the re-emergence energies of different sources according to their cutoff energies. If a source cutoff is at an energy $E_1$ then we expect the re-emergence to be at the energy $E_2$ that gives in (\ref{LIVthreshold}) $\epsilon_{thr}(E_2)=\epsilon_{thr}(E_1)$. Given an ensemble of cutoff and re-emergence energies, we can invoke a systematic statistical procedure, such as Maximum Likelihood, to estimate the most probable value of the LIV parameter, $\xi$. In case the above expectations are realized and the different sources display spectra that fit with high significance a certain LIV scenario, we will have very strong evidence for Lorentz violation, that would otherwise require extremely fine-tuned intrinsic effects or background spectrum. \par

Alternatively, if we ever see an extra-galactic cutoff (or notice an increase in extra-galactic extinction) in a source's gamma-ray spectrum at an energy $E$, this will be a clear sign that a LIV model with $E^*<E$ is not a correct description of nature. We will have [see Eq. (\ref{EStar})] a bound $\xi>[E^{n+2}(n+1)(1-2^{-n})/(4m_e^2)]^\frac{1}{n}/E_{pl}$. In the presence of LIV with symmetry breaking at the Planck scale, the optical depth never increases above $E^*=23$ TeV (note that far infra-red or CMB photons do not make any contribution to the optical depth). We must, of course, be convinced that the break in the spectrum is the result of extra-galactic attenuation and not an intrinsic feature of the source. This will be made possible by observations of cutoffs in the spectra of several different sources. Such observations will help us understand the intrinsic features of blazars, as breaks in the spectra that are not correlated with the sources' distances and appear at energies corresponding to the specifications of the blazars could be interpreted as due to the blazars' mechanism, while breaks at energies that depend linearly on the sources' distances are most likely the result of cosmological absorption. A statistical analysis method can be used here as well to obtain a significant bound on the symmetry breaking scale. \par

The notion of obtaining insight on LIV as described will of course remain a strictly theoretical idea, unless we can make the necessary observations and detect the radiation from the distant gamma-ray sources up to the relevant energies. Luckily, IACTs for very-high-energy gamma-ray astronomy are currently being rapidly enhanced. Two of the most important telescope arrays, MAGIC and HESS, are now being upgraded to Phase-II with the addition of improved larger telescopes. They will have an unprecedented light collection area. This will mainly extend the detection range to lower gamma-ray energies, but a slight increase in the detection rates of multi-TeV photons could, as discussed, be enough to make a significant discovery. A further improvement of the sensitivity in the multi-TeV gamma-ray band is expected with next-generation observatories such as CTA. Only nearby cosmological sources are relevant for the investigation of LIV (more distant sources would be attenuated at energies where LIV is negligible). Luckily again, at the source redshift of $0.03$, demonstrated in Figs. \ref{fig:BlazarSpectrum},\ref{fig:Detection} to be highly sensitive to LIV, there are already two blazars known to be powerful TeV gamma-ray emitters. By now Mkn-501 and Mkn-421 have been observed up to $20$ TeV. With only a modest expectation from the upgraded IACTs, these or other nearby sources can be detected in the near future at slightly higher energies that will necessarily provide us input on the LIV possibility. The more distant gamma-ray sources will serve us for comparison with classical extra-galactic absorption features. \par

Concluding, we have demonstrated a strategy that can test the LIV hypothesis by utilizing its unique signature in observations of cosmological gamma-ray sources. The proposed examination of absorption features is independent of the background photon spectrum. We see that through a simple analysis of the observational data that is expected in the near future, we will either witness for the first time a clear sign of Lorentz violation or have a convincing generic argument against most theories that predict the break of the Lorentz symmetry at the Planck scale.

\begin{acknowledgments}
This research was supported by the ISF Center for Excellence in High Energy Astrophysics.
\end{acknowledgments}

\end{document}